\documentclass[aip,jcp,preprint,groupedaddress]{revtex4-1}
 
\usepackage{graphicx}
\usepackage{dcolumn}
\usepackage{bm}
\usepackage{amsmath}
\usepackage{textcomp}
\usepackage{longtable}
\usepackage{url}
\usepackage{tabularx}
\usepackage{threeparttable}
\usepackage{multirow}
\usepackage{mathtools}

\usepackage[bookmarks=false, hidelinks, linktocpage=true]{hyperref}

\begin{document}

\title{On the crystalline structure of orthorhombic SrRuO$_3$: A benchmark study of DFT functionals}

\author{\v{S}. Masys }
\email[Corresponding author. Email address: ]{Sarunas.Masys@tfai.vu.lt}

\author{V. Jonauskas}

\affiliation{Institute of Theoretical Physics and Astronomy, Vilnius University, A. Go\v{s}tauto Street 12, LT-01108 Vilnius, Lithuania}

\date{\today}

\begin{abstract}
By investigating the crystalline structure of ground-state orthorhombic SrRuO$_3$, we present a benchmark study of some of the most popular density functional theory (DFT) approaches from the local density approximation (LDA), generalized-gradient approximation (GGA), and hybrid functional families. Recent experimental success in stabilizing tetragonal and monoclinic phases of SrRuO$_3$ at room temperature sheds a new light on the ability to accurately describe geometry of this material by applying first-principles calculations. Therefore, our work is aimed to analyse the performance of different DFT functionals and provide some recommendations for future research of SrRuO$_3$. A comparison of the obtained results to the low-temperature experimental data indicates that revised GGAs for solids are the best choice for the lattice constants and volume due to their nice accuracy and low computational cost. However, when tilting and rotation angles appear on the scene, a combination of the revised GGAs with the hybrid scheme becomes the most preferable option. It is important to note that a worse performance of LDA functional is somewhat compensated by its realistic reproduction of electronic and magnetic structure of SrRuO$_3$, making it a strong competitor if the physical features are also taken into account.           
 
\end{abstract}


\maketitle

\section{Introduction}

SrRuO$_3$ is a metallic ferromagnetic perovskite-type ruthenate highly valued for its beneficial electrical and magnetic properties, thermal and chemical stability, atomically smooth surface, and a good lattice match with a wide variety of functional oxides [\onlinecite{koster_68}]. In recent years, SrRuO$_{3}$ has become the most popular epitaxial electrode for complex oxide heterostructures and has been utilized in ferroelectrics, Schottky junctions, magnetocalorics, and magnetoelectrics [\onlinecite{junquera_69, fujii_70, thota_71, niranjan_72}]. Moreover, it has attracted the attention of superconductor and spintronics communities [\onlinecite{takahashi_73, gausepohl_74, feigenson_75}]. Reflecting this broad interest, more than 1000 papers involving the physics, materials science, and applications of SrRuO$_{3}$ have been published over the last two decades [\onlinecite{koster_68}]. But despite all that has been learned, there remain some issues that have not been fully addressed yet.   

At ambient conditions, bulk SrRuO$_3$ crystallizes in an orthorhombic ($Pbnm$) structure and upon heating undergoes phase transitions to tetragonal ($I4/mcm$ at 820 K) and then to cubic ($Pm\bar{3}m$ at 950 K) systems [\onlinecite{kennedy_76}]. Although one can find several previous theoretical studies [\onlinecite{zayak_77, rondinelli_78, hadipour_79, garcia_80, verissimo_81, zang_103, miao_82}] in which the ground-state orthorhombic structure is reproduced by density functional theory (DFT) calculations within the local density approximation (LDA), generalized-gradient approximation (GGA), hybrid functional, and even $\text{LDA}+U$ or $\text{GGA}+U$ frameworks, a more detailed and systematic analysis of the obtained results in most cases is missing. This now seems like an obvious gap of knowledge, since the importance of geometry optimization has recently grown to a new level due to the experimental breakthrough in stabilizing tetragonal and/or monoclinic phases of SrRuO$_3$ at room temperature by applying external pressure [\onlinecite{zhernenkov_84}], introducing oxygen vacancies [\onlinecite{lu_85}, \onlinecite{lu_86}], and employing compressive and/or tensile strains induced by different substrates [\onlinecite{vailionis_87, choi_88, chang_89, vailionis_90, herklotz_91, kan_92, kan_93, aso_94, lu_95}]. As it was noticed by Herklotz $et$ $al$. [\onlinecite{herklotz_96}], an ongoing debate on the exact space group symmetry of the strained SrRuO$_3$ films paves the way for the first-principles calculations which can lend a helping hand to the researchers in the laboratory. For example, in the work of Vailionis $et$ $al$. [\onlinecite{vailionis_90}] the authors state that a tensile strained SrRuO$_3$ thin film most likely possesses a $Cmcm$ rather than $I4/mmm$ space group symmetry, but the subtle differences between these two phases are too small to be detected experimentally. From a theoretical standpoint, this issue could be more or less painlessly resolved by carefully performing geometry optimization at the DFT level. What is more, geometry optimization also plays a significant role in theoretical modelling the electronic structure of SrRuO$_3$ films grown on SrTiO$_3$ substrates at the ultrathin limit [\onlinecite{mahadevan_97, gupta_98, liang_99}]. However, in order to effectively employ DFT approaches for the strained systems, initially one should get acquainted with their side-by-side performance comparison produced for the ground-state configuration. 

In this study, we have systematically investigated the $Pbnm$ crystalline structure of SrRuO$_3$ and compared obtained results with the low-temperature experimental data. From a theoretical point of view, we have comprised three rungs of the Jacob's ladder of DFT approximations [\onlinecite{perdew_100}] -- the first, second, and fourth -- by applying LDA, GGAs, and hybrid functionals that are widely used for the solid-state calculations. The third rung of the ladder assigned for the meta-GGAs, which in turn consider kinetic-energy density as an additional variable, was not taken into account, since promising meta-GGAs for solids, like revTPSS [\onlinecite{perdew_101}], are still not implemented in many of the popular DFT codes. We have distinguished structural parameters into three categories, namely, (a) lattice constants and volume, (b) tilting and rotation angles, and (c) internal angles and bond distances within RuO$_6$ octahedra. On one hand, it is well known that lattice constants and volume are critically important for many inherent material properties -- including phonon frequencies, elastic constants, ferromagnetism, and the possibility of structural phase transitions [\onlinecite{csonka_102}, \onlinecite{haas_2010_22}] -- and therefore deserve an exceptional attention. On the other hand, it has been shown that octahedral rotations and tilts in the strained SrRuO$_3$ thin films are a key factor for determining their functionalities and possible applications [\onlinecite{lu_86}, \onlinecite{vailionis_90}, \onlinecite{kan_92}, \onlinecite{lu_95}]. Experimental recognition of epitaxially stabilized tetragonal structures with no octahedral rotations, but with octahedral deformations introduced instead [\onlinecite{kan_93}, \onlinecite{aso_94}] encourages to focus on the internal parameters of octahedra. As the electrical and magnetic properties of these thin films were discovered to exhibit a close relationship with the strain-induced octahedral deformations, we also found it important to check the performance of DFT approaches in reproducing the ground-state geometry of RuO$_6$. For the sake of completeness, we have additionally presented the electronic structure and magnetic moment of SrRuO$_3$ which allow us to demonstrate the versatility of the tested functionals in the context of physical features.    

\section{Theoretical background}    

In the LDA framework [\onlinecite{lda_56}], the exchange energy $E_{\text{X}}[n]$ has the form
\begin{equation}
E_{\text{X}}^{\text{LDA}}[n]=\int n\varepsilon_{\text{X}}^{\text{LDA}}(n)d^{3}r,
\label{eq:equ1}
\end{equation}
where $n$ denotes the electron density and $\varepsilon_{\text{X}}^{\text{LDA}}(n)=-(3/4)(3/\pi)^{1/3}n^{1/3}$ is the exchange energy density per particle for a uniform electron gas. The GGA form for the exchange energy is simply
\begin{equation}
E_{\text{X}}^{\text{GGA}}[n]=\int n\varepsilon_{\text{X}}^{\text{LDA}}(n)F_{\text{X}}(s)d^{3}r,
\label{eq:equ2}
\end{equation}
in which $s=\vert\nabla n\vert/(2k_{\text{F}}n)$ (with Fermi wave vector $k_{\text{F}}=(3\pi^{2}n)^{1/3}$) is the dimensionless reduced gradient and $F_{\text{X}}(s)$ is the exchange enhancement factor. Any GGA that reproduces the uniform gas limit can be expressed as [\onlinecite{pbesol_18}]
\begin{equation}
F_{\text{X}}(s)=1+\mu_{\text{GE}}s^{2}+\ldots (s\rightarrow 0),
\label{eq:equ3}
\end{equation}
and accordingly
\begin{equation}
E_{\text{X}}^{\text{GGA}}[n]=\int n\varepsilon_{\text{X}}^{\text{LDA}}(n)\lbrace 1+\mu_{\text{GE}}s^{2}+\ldots \rbrace d^{3}r=E_{\text{X}}^{\text{LDA}}[n]+\int n\varepsilon_{\text{X}}^{\text{LDA}}(n)\lbrace \mu_{\text{GE}}s^{2}+\ldots \rbrace d^{3}r,
\label{eq:equ4}
\end{equation}
where the gradient expansion (GE) that is precise for slowly-varying electron gases has [\onlinecite{antono_1985_39}]
\begin{equation}
\mu_{\text{GE}}=10/81\approx0.1235 .
\label{eq:equ5}
\end{equation}
The PBE GGA [\onlinecite{pbe_57}] is nowadays considered as a standard functional for solid-state calculations [\onlinecite{haas_tran_2009_40}]. Although this GGA belongs to the class of parameter-free functionals, it still contains some arbitrary choices, e.g., the analytical form of the enhancement factor or the constraints that have to be satisfied. It has been shown [\onlinecite{pbesol_18}, \onlinecite{perdew_2006_41}] that in order to obtain the accurate exchange energy for free neutral atoms, any GGA must have $\mu\approx2\mu_{\text{GE}}$. For the PBE functional, $\mu$ is set to 0.2195 from slightly different requirement which is based on the reproduction of the LDA jellium response. Here, $F_{\text{X}}(s)$ has the form
\begin{equation}
F_{\text{X}}^{\text{PBE}}(s)=1+\kappa\left(1-\frac{1}{1+\frac{\mu s^{2}}{\kappa}}\right).
\label{eq:equ6}
\end{equation}
The parameter $\kappa$, which controls behaviour at $s \rightarrow \infty$, is set to 0.804 according to the relation $\kappa=\lambda_{\text{LO}}/2^{1/3}-1$ to ensure the Lieb-Oxford (LO) bound [\onlinecite{lieb_1981_42}], which is an upper limit on the ratio of the exact exchange-correlation energy to the value of the LDA approximation of the exchange energy $(E_{\text{X}}[n]\geq E_{\text{XC}}[n]\geq \lambda_{\text{LO}}E_{\text{X}}^{\text{LDA}}[n]$ with $\lambda_{\text{LO}}=2.273)$. At the limit of a slowly-varying high density, the GE for the correlation energy of a GGA can be written as
\begin{equation}
E_{\text{C}}^{\text{GGA}}[n]=\int n\lbrace \varepsilon_{\text{C}}^{\text{LDA}}(n)+\beta_{\text{GE}}t^{2}+\ldots \rbrace d^{3}r=E_{\text{C}}^{\text{LDA}}[n]+\int n\lbrace \beta_{\text{GE}}t^{2}+\ldots \rbrace d^{3}r,
\label{eq:equ7}
\end{equation}
where $\varepsilon_{\text{C}}^{\text{LDA}}(n)$ is the correlation energy per particle of the uniform electron gas, $\beta_{\text{GE}}$ is a coefficient set to 0.0667 [\onlinecite{ma_1968_43}], and $t=\vert\nabla n\vert/(2k_{\text{TF}}n)$ (with Thomas-Fermi screening wave vector $k_{\text{TF}}=\sqrt{4k_{\text{F}}/\pi}$) denotes an appropriate reduced density gradient for correlation. In the PBE correlation functional, the value of $\beta_{\text{GE}}$ is retained, whereas in PBEsol [\onlinecite{pbesol_18}], it is chosen to be 0.046 in order to reproduce the accurate exchange-correlation energy for a jellium surface obtained at the meta-GGA TPSS [\onlinecite{tao_2003_44}] level. In the PBEsol exchange functional, the value of $\mu$, which determines behaviour for $s\rightarrow 0$, is restored back to $\mu_{\text{GE}}=10/81$, since it has been argued [\onlinecite{pbesol_18}] that $\mu\approx2\mu_{\text{GE}}$ is harmful for many condensed matter applications. This choice allows to recover the second-order GE, but on the other hand, it means that PBEsol no longer satisfies the LDA jellium response, because $\mu\neq \pi^{2}\beta/3$. Thus, as $s\rightarrow 0$, there is no complete cancellation between beyond-LDA exchange and correlation contributions. While constructing the PBE functional, this sort of cancellation was believed to be more accurate than the lower-order gradient expansion for small $s$. Although PBE works equally well for finite and infinite systems [\onlinecite{haas_tran_2009_40}], PBEsol outperforms it in various crystalline structure calculations. Nevertheless, this benefit is accompanied by a worsening of the thermochemical properties [\onlinecite{sogga_19}]. It is evident that due to pretty simple mathematical form of GGA one has to choose between improved atomization energies of molecules or improved lattice parameters of solids [\onlinecite{haas_2010_22}].    

The SOGGA exchange functional [\onlinecite{sogga_19}], used in combination with the PBE correlation functional, completely restores the GE to the second order for both exchange and correlation. The analytical form of the SOGGA exchange enhancement factor is expressed as an average of the PBE and RPBE [\onlinecite{hammer_1999_45}] exchange functionals
\begin{equation}
F_{\text{X}}^{\text{SOGGA}}(s)=1+\kappa\left(1-\frac{1}{2}\cdot \frac{1}{1+\frac{\mu s^{2}}{\kappa}}-\frac{1}{2}\cdot \text{exp}\left(-\frac{\mu s^{2}}{\kappa}\right)\right),
\label{eq:equ8}
\end{equation} 
in which $\mu=\mu_{\text{GE}}=10/81$. The parameter $\kappa$ is set to 0.552 in order to satisfy a tighter LO bound $(E_{\text{X}}[n]\geq E_{\text{XC}}[n]\geq \lambda_{\text{tLO}}E_{\text{X}}^{\text{LDA}}[n]$ with $\lambda_{\text{tLO}}=1.9555)$. 

The WC exchange enhancement factor is given by [\onlinecite{wc_20}]
\begin{equation}
F_{\text{X}}^{\text{WC}}(s)=1+\kappa\left(1-\frac{1}{1+\frac{x(s)}{\kappa}}\right),
\label{eq:equ9}
\end{equation} 
where
\begin{equation}
x(s)=\underbrace{\frac{10}{81}s^{2}}_\text{$1^{\text{st}}$ term}+\underbrace{\left(\mu - \frac{10}{81} \right)s^{2}\text{exp}\left(-s^{2}\right)}_\text{$2^{\text{nd}}$ term}+\underbrace{\text{ln}\left(1+cs^{4}\right)}_\text{$3^{\text{rd}}$ term}.
\label{eq:equ10}
\end{equation} 
Parameters $\kappa$ and $\mu$ have the same values as in PBE and $c=0.0079325$ is set to recover the fourth order parameters of the fourth order GE of the exchange functional for small $s$ (unfortunately, incorrectly [\onlinecite{haas_tran_2009_40}]). On the whole, the analysis of a large set of solids [\onlinecite{haas_2009_24}] shows that concerning the lattice constants PBEsol, SOGGA, and WC perform quite similarly, and in most cases demonstrate an explicit improvement over the typically underestimated and overestimated values of LDA and PBE, respectively. 
 
It is interesting to note that for some solids the overestimated lattice constants can be improved by including a fraction of the exact Hartree-Fock (HF) exchange energy [\onlinecite{paier_45, b3lyp_fail, marsman_46}]. Among the so called hybrid functionals, PBE0 [\onlinecite{pbe0_47}] is defined by
\begin{equation}
E_{\text{XC}}^{\text{PBE0}}[n]=aE_{\text{X}}^{\text{HF}}+(1-a)E_{\text{X}}^{\text{PBE}}[n]+E_{\text{C}}^{\text{PBE}}[n],
\label{eq:equ11}
\end{equation} 
where the amount of mixing coefficient $a=1/4$ has been derived from theoretical arguments through the perturbation theory. However, due to the slow decay of the HF exchange interaction, which in some cases can reach up to hundreds of angstroms, the evaluation of $E_{\text{X}}^{\text{HF}}$ may be computationally very demanding. Fortunately, it was demonstrated [\onlinecite{heyd_48}] that the screened HF exchange, in which the computationally expensive long-range part is replaced by a corresponding PBE counterpart, exhibits all physically relevant properties of the full HF exchange. The resulting expression for the exchange-correlation energy, known as HSE06 [\onlinecite{krukau_49}], is given by
\begin{equation}
E_{\text{XC}}^{\text{HSE06}}[n]=aE_{\text{X}}^{\text{HF,SR},\omega}+(1-a)E_{\text{X}}^{\text{PBE,SR},\omega}[n]+E_{\text{X}}^{\text{PBE,LR},\omega}[n]+E_{\text{C}}^{\text{PBE}}[n].
\label{eq:equ12}
\end{equation} 
Here, (SR) and (LR) denote the short- and long-range parts of the respective exchange interactions. The separation of them is achieved through a partitioning of the Coulomb potential for exchange
\begin{equation}
\frac{1}{r}=\underbrace{\frac{\text{erfc}(\omega r)}{r}}_\text{SR}+\underbrace{\frac{\text{erf}(\omega r)}{r}}_\text{LR},
\label{eq:equ13}
\end{equation} 
where $r = \vert \mathbf{r}-\mathbf{r'}\vert$, and $\omega$ is the screening parameter that defines the separation range. For $\omega=0$, the PBE0 functional is recovered, and for $\omega\rightarrow \infty$, HSE06 becomes identical to PBE. The value of the screening parameter $\omega=0.11$ bohr$^{-1}$ was chosen as a reasonable compromise between computational cost and the quality of the results which, regarding the lattice constants, are similar to those obtained using the PBE0 functional [\onlinecite{marsman_46}]. 

Interestingly, it appears that the accuracy of the lattice constants can be increased for both screened and non-screened hybrid schemes by substituting the PBE functional with the ones revised for solids. For the screened hybrids, the HSEsol functional [\onlinecite{hsesol_50}] was introduced which has the same form (see Eq. \ref{eq:equ12}) and the same range-separation parameter as HSE06, but it is based on PBEsol for the exchange as well as correlation part. Concerning the non-screened hybrids, the B1WC functional [\onlinecite{b1wc_51}] -- which is in turn based on the PBE0 scheme (see Eq. \ref{eq:equ11}) with the PBE exchange and correlation correspondingly replaced by WC and PW [\onlinecite{pw_52}], and parameter $a$ set to 0.16 -- also shows promising results, especially for the perovskite-structured materials [\onlinecite{garcia_53}].   

\section{Computational details}

The ferromagnetic state of low-temperature orthorhombic SrRuO$_{3}$ was modelled with the CRYSTAL14 code [\onlinecite{crystal14_54}] employing a linear combination of atom-centered Gaussian orbitals. The small-core Hay-Wadt pseudopotentials [\onlinecite{hay_59}] were utilized to describe the inner-shell electrons ($1s^{2}2s^{2}2p^{6}3s^{2}3p^{6}3d^{10}$) of Sr and Ru atoms. The valence part of the basis set for Sr ($4s^{2}4p^{6}5s^{2}$) was adopted from the SrTiO$_{3}$ study [\onlinecite{piskunov_60}], while the valence functions for Ru ($4s^{2}4p^{6}4d^{7}5s^{1}$) were taken from our previous work on non-stoichiometric SrRuO$_{3}$ [\onlinecite{masys_61}]. Regarding the O atom, the all-electron basis set with a double set of $d$ functions was applied from the CaCO$_{3}$ study [\onlinecite{valenzano_62}]. All these aforementioned basis sets are also available online at CRYSTAL's basis sets library [\onlinecite{crystal_63}].     

While performing full geometry optimization, the default values were chosen for most of the technical setup, the details of which can be found in CRYSTAL14 user's manual [\onlinecite{crystal14_55}]. However, the parameters that define the convergence threshold on total energy and truncation criteria for bielectronic integrals were tightened to (8) and (8 8 8 8 16), respectively. Besides, the allowed root-mean-square values of energy gradients and nuclear displacements were correspondingly set to more severe ones: 0.00006 and 0.00012 in atomic units. In order to improve the self-consistence field convergence, the Kohn-Sham matrix mixing technique (at $80\%$) together with Anderson's method [\onlinecite{anderson_64}], as proposed by Hamman [\onlinecite{hamann_65}], were applied. The reciprocal space was sampled according to regular sublattice with a shrinking factor of 8, resulting in 125 independent \textbf{\textit{k}} points in the first irreducible Brillouin zone.

We would like to remark that in this paper instead of the original PW functional the correlation part from the PBE framework was taken for the modified B1WC calculations, named as mB1WC. In the meantime, LDA exchange was combined with VWN correlation [\onlinecite{lda_58}], whereas PBE, SOGGA, and WC exchange functionals were used with the correlation part of PBE. The PBEsol exchange functional was employed with the correlation part of PBEsol and PBE. A separate notation PBEsol\textsuperscript{PBE} was introduced for the latter combination.

\section{Results and discussion}

\begin{figure}
\centering
{
\includegraphics[scale=0.75]{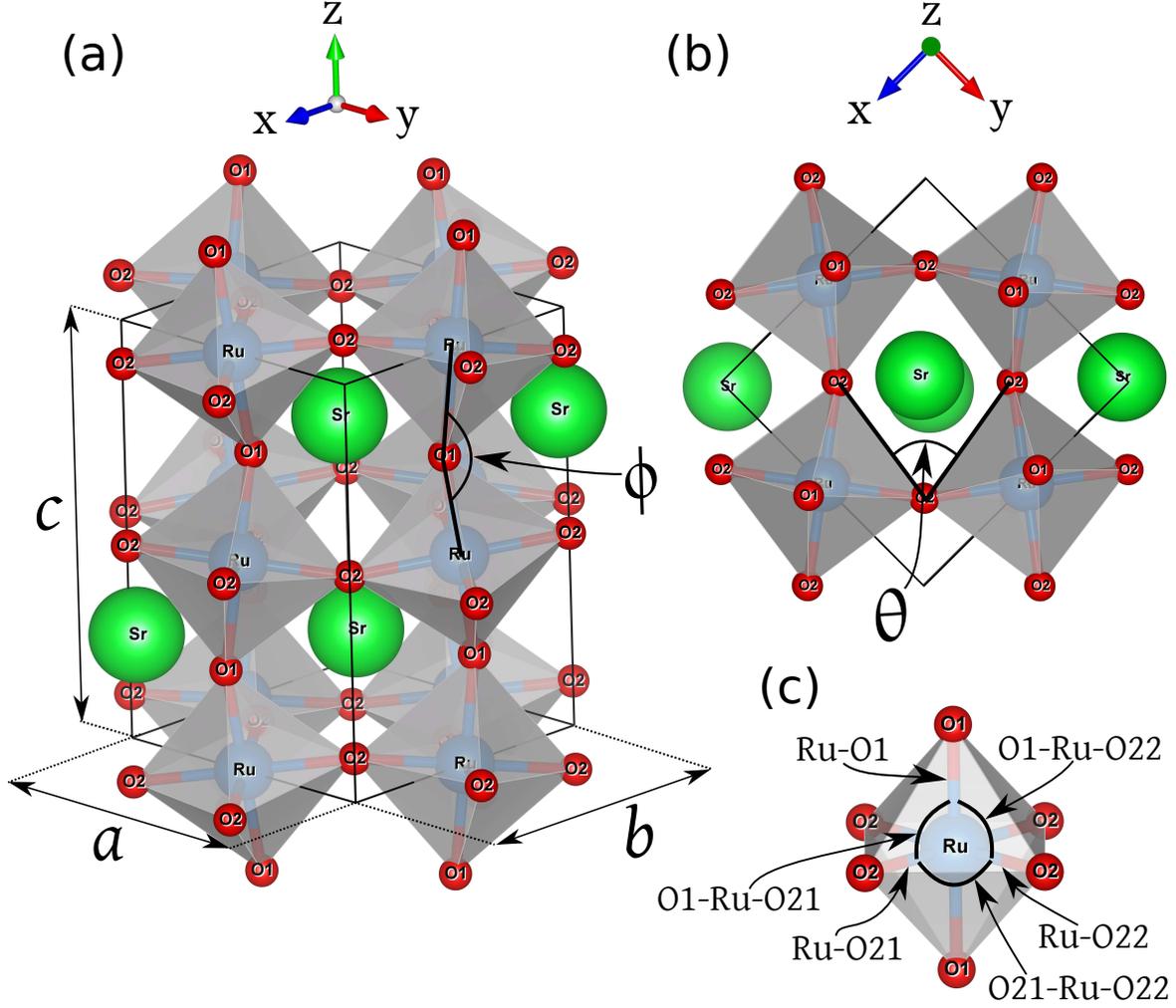}
}
\caption{\label{fig1}
Schematic representation of (a) the crystalline structure of ($Pbnm$) SrRuO$_3$, (b) its top view, and (c) octahedral parameters. Notation O1 and O2 labels oxygen atoms at the apical and planar positions of the RuO$_6$ octahedra, respectively. The drawings were produced with the visualization program VESTA [\onlinecite{vesta_67}].}
\end{figure}

For the sake of clarity, the geometry of ground-state SrRuO$_3$ is visualized in Fig. \ref{fig1}. The equilibrium structural parameters and magnetic moment calculated within different DFT approximations are given in Table \ref{tab1}. The low-temperature experimental data are also presented therein, however, no zero-point anharmonic expansion (ZPAE) corrections were applied. On one hand, it is not straightforward to evaluate ZPAE corrections for the non-cubic systems. On the other hand, ZPAE should not have a noticeable influence on material like SrRuO$_3$. Our previous non-magnetic calculations [\onlinecite{masys_104}] indicate that ZPAE correction for the lattice constant of cubic SrRuO$_3$ reaches at most $\sim0.13\%$ and therefore can be treated as negligible. The given mean absolute relative errors (MAREs) were evaluated according to the expression
\begin{equation}
\text{MARE} = \frac{100}{n}\displaystyle\sum\limits_{i=1}^{n}{\bigg \vert \frac{p_{i}^{\text{Calc.}}-p_{i}^{\text{Expt.}}}{p_{i}^{\text{Expt.}}} \bigg \vert},
\label{eq:equ14}
\end{equation}
in which $p_{i}^{\text{Calc.}}$ and $p_{i}^{\text{Expt.}}$ are the calculated and experimental values of the considered parameter, respectively. Note that in the literature tilting and rotation angles of RuO$_6$ octahedra are usually defined through the corresponding relations with $\phi$ and $\theta$:
\begin{equation}
\Phi=\frac{\left(180^{\circ}-\phi \right)}{2},
\label{eq:equ15}
\end{equation}
\begin{equation}
\Theta=\frac{\left(90^{\circ}-\theta \right)}{2}.
\label{eq:equ16}
\end{equation}  
But in this study, $\phi$ and $\theta$ are themselves denoted as tilting and rotation angles, since -- as it can be seen from Fig. \ref{fig1} -- they can be measured directly. The derived values of $\Phi$ and $\Theta$ would simply distort the provided statistics in regard to other structural parameters. 

\begin{table}
\caption{\label{tab1}
Calculated structural parameters and magnetic moment of orthorhombic ($Pbnm$) SrRuO$_{3}$ compared to the experimental data at 1.5 K and 10 K [\onlinecite{bushmeleva_66}], respectively. Lattice constants $a$, $b$, and $c$ together with bond distances Ru-O1, Ru-O21, and Ru-O22 are given in \AA, volume $V$ is given in \AA$^{3}$, angles $\phi$, $\theta$, O1-Ru-O21, O1-Ru-O22, and O21-Ru-O22 are given in degrees, magnetic moment $\mu$ is given in $\mu_{\text{B}}$ per formula unit. MARE (in $\%$) stands for the mean absolute relative error: MARE$_1$ is evaluated for $a$, $b$, $c$, and $V$; MARE$_2$ for $\phi$ and $\theta$; MARE$_3$ for Ru-O1, Ru-O21, Ru-O22, O1-Ru-O21, O1-Ru-O22, and O21-Ru-O22; MARE$_\text{T}$ denotes the total MARE of all 12 structural parameters. The numbers in brackets (in $\%$) represent absolute relative errors for each structural parameter. Magnetic moment is taken from the calculations with the experimental geometry at 1.5 K.      }
\scriptsize
\begin{tabularx}{\textwidth}{>{\hsize=1.5\hsize}X>{\centering\arraybackslash\hsize=0.9\hsize}X>{\centering\arraybackslash\hsize=0.9\hsize}X>{\centering\arraybackslash\hsize=0.94\hsize}X>{\centering\arraybackslash\hsize=1.26\hsize}X
>{\centering\arraybackslash\hsize=0.94\hsize}X>{\centering\arraybackslash\hsize=0.9\hsize}X>{\centering\arraybackslash\hsize=0.9\hsize}X>{\centering\arraybackslash\hsize=0.9\hsize}X>{\centering\arraybackslash\hsize=0.9\hsize}X>{\centering\arraybackslash\hsize=0.9\hsize}X>{\centering\arraybackslash\hsize=0.9\hsize}X}
 \hline \hline  
                   &  LDA    &   PBE   & PBEsol & PBEsol\textsuperscript{PBE} &   SOGGA  &   WC    & mB1WC    &  PBE0    & HSE06   &  HSEsol  & Expt.    \\ 
\hline
$a$                &   5.530 &  5.644  &  5.577 &   5.568                     &  5.565   &   5.580 &    5.555 &   5.557  &   5.564 &   5.534  &  5.566   \\
                   &   (0.64)&  (1.40) &  (0.20)&   (0.04)                    &  (0.01)  & (0.26)  &   (0.20) &  (0.17)  &   (0.03)&   (0.58) &          \\
$b$                &   5.490 &  5.623  &  5.549 &   5.538                     &  5.534   &   5.553 &  5.512   &  5.580   &  5.572  &   5.503  &   5.532  \\
                   &   (0.76)&  (1.65) & (0.31) &  (0.10)                     &   (0.04) & (0.38)  &  (0.36)  &   (0.86) &   (0.72)&   (0.52) &          \\
$c$                &   7.787 &  7.970  &  7.871 &   7.858                     &  7.854   &   7.877 &  7.831   &  7.915   &  7.915  &   7.839  &   7.845  \\
                   &   (0.74)&  (1.59) & (0.33) &  (0.16)                     &   (0.11) & (0.41)  &  (0.18)  &   (0.89) &  (0.90) &   (0.07) &          \\
$V$                &  236.41 &  252.94 & 243.58 &  242.30                     &  241.90  &  244.08 &  239.78  &  245.39  &  245.40 &   238.75 &   241.56 \\
                   &   (2.13)& (4.71)  & (0.84) &  (0.31)                     &  (0.14)  & (1.04)  &  (0.73)  &   (1.58) &  (1.59) &   (1.16) &          \\
$\phi$             &   159.31&  159.26 &  159.76&   159.98                    &  159.97  &  159.58 &  161.24  &  159.10  &  159.71 &   161.09 &   161.99 \\
                   &   (1.66)&  (1.69) & (1.38) &  (1.24)                     &   (1.25) & (1.49)  &  (0.46)  &   (1.78) &  (1.41) &   (0.56) &          \\
$\theta$           &  75.08  &  74.20  &  74.28 &   74.43                     &  74.40   &   74.21 &  76.20   &  74.73   & 75.17   &   76.18  &   77.31  \\
                   &   (2.88)&  (4.03) & (3.92) &  (3.73)                     &   (3.76) & (4.01)  &  (1.43)  &   (3.34) &  (2.78) &   (1.46) &          \\
Ru-O1              &   1.979 &  2.026  &  1.999 &  1.995                      &  1.994   &   2.001 &  1.984   &  2.012   &   2.010 &   1.987  &   1.986  \\
                   &   (0.34)&  (2.00) & (0.65) &  (0.45)                     &  (0.40)  & (0.76)  &   (0.08) &  (1.32)  &  (1.23) &   (0.06) &          \\
Ru-O21             &   1.983 &  2.030  &  2.004 &  1.999                      &  1.998   &   2.005 &  1.986   &  1.979   &    1.979&  1.961   &   1.986  \\
                   &   (0.13)&  (2.24) & (0.89) &  (0.66)                     &  (0.61)  & (0.98)  &   (0.02) &  (0.36)  &  (0.33) &   (1.24) &          \\  
Ru-O22             &   1.981 &  2.028  &  2.001 &  1.997                      &  1.996   &   2.003 &  1.984   &  2.023   &    2.020&   1.997  &   1.987  \\
                   &   (0.30)&  (2.03) & (0.70) &  (0.49)                     &  (0.44)  & (0.79)  &   (0.16) &  (1.81)  &  (1.65) &   (0.48) &          \\
O1-Ru-O21          &  90.20  &  90.04  & 90.16  &  90.20                      & 90.21    &  90.14  &   90.15  & 90.99    &   90.88 &  90.69   &   90.33  \\
                   &  (0.15) &  (0.32) & (0.20) &  (0.15)                     & (0.14)   & (0.22)  &    (0.20)&  (0.72)  & (0.61)  &   (0.39) &          \\
O1-Ru-O22          &   90.46 &  90.25  &  90.33 &   90.38                     &  90.39   &   90.33 &  90.39   &  89.83   &   90.03 &   90.51  &   90.27  \\
                   & (0.21)  &  (0.02) & (0.06) &  (0.12)                     & (0.13)   & (0.07)  &   (0.14) &   (0.49) &   (0.27)&   (0.26) &          \\
O21-Ru-O22         &   91.38 &  91.17  &  91.19 &   91.21                     &  91.21   &   91.21 &  91.24   &  90.75   &   90.87 &   91.21  &   91.07  \\
                   & (0.35)  &  (0.12) & (0.14) &  (0.16)                     & (0.16)   & (0.16)  &   (0.19) &   (0.35) &   (0.21)&   (0.15) &          \\
$\mu$              &    1.70 &  1.96   & 1.92   &  1.93                       &   1.93   &   1.94  &   2.00   &   2.00   &   2.00  &   2.00   &  1.63    \\
MARE$_1$           &    1.07 &  2.34   & 0.42   &  0.15                       &   0.08   &   0.52  &   0.37   &   0.88   &   0.81  &   0.58   &          \\
MARE$_2$           &    2.27 &  2.86   & 2.65   &  2.49                       &   2.50   &   2.75  &   0.95   &   2.56   &   2.09  &   1.01   &          \\ 
MARE$_3$           &    0.25 &  1.12   & 0.44   &  0.34                       &   0.31   &   0.50  &   0.13   &   0.84   &   0.72  &   0.43   &          \\ 
MARE$_\text{T}$    &    0.86 &  1.82   & 0.80   &  0.64                       &   0.60   &   0.88  &   0.34   &   1.14   &   0.98  &   0.58   &          \\    

\hline \hline
\end{tabularx}
\end{table}  

\subsection{Lattice constants and volume}

We start with the lattice constants and volume, and a careful look at Table \ref{tab1} reveals that LDA and PBE functionals tend to underestimate and overestimate these parameters, respectively, with LDA being substantially closer to the experiment than PBE. The revised functionals for solids -- PBEsol, PBEsol\textsuperscript{PBE}, SOGGA, and WC -- demonstrate an explicit improvement over the results of LDA and PBE. That is not very surprising, though, since the same tendency was already noticed for a large set of solids [\onlinecite{haas_2009_24}]. Among the revised functionals, PBEsol\textsuperscript{PBE} and SOGGA show the best performance with the corresponding MARE$_1$ values of 0.15 and $0.08\%$. Having in mind that ``good" theoretical deviations should not exceed $0.5\%$ [\onlinecite{haas_2009_24}], these numbers appear to be truly impressive. The numbers of PBEsol ($0.42\%$) and WC ($0.52\%$) can also be considered as good, while LDA ($1.07\%$) and PBE ($2.34\%$) exceed the critical value of what we call a satisfactory threshold, in this work set to $1\%$. Interestingly, the inclusion of $25\%$ of the exact HF exchange -- as it is in the PBE0 scheme -- has a noticeable influence on the performance of PBE functional, since MARE$_1$ now gets reduced from unacceptable $2.34\%$ to a satisfactory value of $0.88\%$. The additional rejection of the long-range HF exchange in the HSE06 approximation has a small but positive effect, with MARE$_1$ being further improved to $0.81\%$. However, the combination of HSE-type framework and PBEsol-type revision seems to be the best choice among hybrids (HSEsol with $0.58\%$), unless the amount of full-range HF exchange is reduced to $16\%$ and PBEsol functional is interchanged with WC counterpart (mB1WC with $0.37\%$). Notice that these two hybrids, together with LDA, also have some potential to benefit from ZPAE corrections, as their lattice constants and volume are slightly lower compared to experimental data. In the meantime, all the remaining functionals practically have no room for improvement, with their structural parameters being higher than experimental ones. A further comparison to the results of $\text{LDA}+U$ and $\text{GGA}+U$ schemes, presented in the corresponding works of Verissimo-Alves $et$ $al.$ [\onlinecite{verissimo_81}] and Zang $et$ $al.$ [\onlinecite{zang_103}], indicates that the introduction of electron correlation corrections to the Ru $4d$ orbitals has a minor influence on the performance of their respective parent functionals, LDA and PBE, since MARE$_1$ for $\text{LDA}+U$ reaches $1.2\%$, whereas for $\text{GGA}+U$ $2.05\%$. These numbers are very close to our LDA and PBE values, besides, the typical behaviour to underestimate and overestimate lattice parameters also remains unaffected. It appears that the inclusion of the Hubbard $U$ term is not an effective option for a better description of SrRuO$_3$ geometry, at least for the considered values of $U$.     

\subsection{Tilting and rotation angles}

As it was already noted by Garc\'{i}a-Fern\'{a}ndez $et$ $al.$ [\onlinecite{garcia_53}] for a variety of perovskites, calculation of tilting and rotation angles involves energy changes that are much more subtle than those associated with lattice parameters, therefore relative errors of DFT approaches become more pronounced. And indeed, Table \ref{tab1} clearly demonstrates that deviations from experiment are much larger than previously discussed ones, especially for the rotation angle $\theta$. Revisions made to PBEsol, SOGGA, and WC approximations have a quite small positive impact on the performance of PBE and, what is more surprising, are sufficient to outperform LDA only for the tilting angle $\phi$. The results of revised functionals fall within the inner range between MARE$_2$ values of LDA ($2.27\%$) and PBE ($2.86\%$), which are far above our satisfactory threshold. It is worth mentioning that PBE0 ($2.56\%$) and HSE06 ($2.09\%$) hybrids do not essentially ameliorate the situation, although in this case the rejection of the long-range HF exchange shows slightly more pronounced influence compared to the HSE06 numbers for lattice constants and volume. However, the most interesting observation is that neither enhancement factor revision nor inclusion of a portion of exact exchange is able to alone improve the results of PBE to an acceptable level. Only a combination of both has an effect large enough to reduce MARE$_2$ to values near a satisfactory threshold, like $1.01\%$ for HSEsol and $0.95\%$ for mB1WC hybrids. This is a somewhat unexpected trend, once again confirming that variations in tilting and rotation angles are extremely energetically sensitive and difficult to accurately reproduce. An additional comparison to the $\text{LDA}+U$ [\onlinecite{verissimo_81}] performance with its MARE$_2$ reaching $2.38\%$ indicates that a Hubbard-type electron correlation correction is negligible, as we have already remarked from the analysis of lattice parameters.     

\subsection{Internal geometry of octahedra}

In contrast to the tilting and rotation angles, the internal parameters of RuO$_6$ octahedra are reproduced with commendable accuracy. Even the PBE functional -- the worst performer so far -- yields reasonable results with its MARE$_3$ reaching $1.12\%$, which is just slightly above the satisfactory threshold. The rest of the GGAs, on the other hand, do not exceed $0.5\%$ for MARE$_3$ but their good execution remains a bit worse compared to that of LDA ($0.25\%$), especially for the bond distances Ru-O. However, the same cannot be said about the results of $\text{LDA}+U$ [\onlinecite{verissimo_81}] framework, despite the fact that our previous analysis of lattice parameters and interoctahedral angles revealed only a small difference between LDA and $\text{LDA}+U$ performance. In this case, MARE$_3$ for $\text{LDA}+U$ increases to $0.88\%$, and that is the most prominent discrepancy noticed for these two methods. It is no longer surprising that the inclusion of a portion of HF exchange has a positive impact on the results of PBE with MARE$_3$ being reduced to $0.84\%$ for PBE0 and $0.72\%$ for HSE06. Although the interchange with PBEsol functional gives even a better effect (HSEsol with $0.43\%$), the closest resemblance to the experiment was demonstrated by mB1WC hybrid with its excellent MARE$_3$ value of $0.13\%$. One can note that mB1WC is able to outperform LDA for practically every parameter of RuO$_6$. On the whole, the performance of all presented functionals can be considered as rather good and, to be honest, that is not a very astonishing trend. Octahedral deformations in SrRuO$_3$ are very energetically costly [\onlinecite{lu_95}], therefore large energy variations involved in RuO$_6$ geometry optimization are easier to deal with for a variety of DFT approximations.

\subsection{Overall performance}

The values of MARE$_\text{T}$ reflect the overall performance of tested functionals in reproducing the ground-state crystalline structure of SrRuO$_3$ by taking into account both the internal geometry of the unit cell and lattice parameters. From Table \ref{tab1} it is seen that despite its well-known versatility for finite and infinite systems the original PBE approach undoubtedly remains the last selection among all the contestants. Its MARE$_\text{T}$ value of $1.82\%$ practically two times exceeds our satisfactory threshold, and even the incorporation of $25\%$ of the exact full-range HF exchange does not help to achieve desired MARE$_{\text{T}} < 1\%$ (PBE0 with $1.14\%$). On the other hand, the screening of HF exchange in the HSE-type framework is effective enough to improve MARE$_\text{T}$ to a satisfactory result (HSE06 with $0.98\%$), although the difference between the performance of these two hybrids is not very significant. The more surprising result is demonstrated by LDA functional, since its value of $0.86\%$ is not only noticeably smaller than that of $\text{LDA}+U$ [\onlinecite{verissimo_81}] scheme ($1.24\%$) -- conclusively convincing that the Hubbard $U$ term is unnecessary for the geometry optimization of bulk SrRuO$_3$, -- but is also more or less of the same magnitude as MARE$_\text{T}$ of revised WC ($0.88\%$) and PBEsol ($0.8\%$) approximations. The initial advantage of these two GGAs obtained while reproducing lattice constants and volume is gradually lost when internal geometry of SrRuO$_3$ is taken into consideration. And yet, the combination of $E_{\text{X}}^{\text{PBEsol}}[n]$ and $E_{\text{C}}^{\text{PBE}}[n]$ as well as modifications present in $F_{\text{X}}^{\text{SOGGA}}(s)$ appear to be sufficiently fruitful to bring respective MARE$_\text{T}$ of PBEsol\textsuperscript{PBE} ($0.64\%$) and SOGGA ($0.6\%$) close to ``good" theoretical values ($0.5\%$). A similar result is also yielded by HSEsol hybrid ($0.58\%$), but in this case the largest benefit is derived from the superior reproduction of tilting and rotation angles. However, a MARE$_\text{T}$ value of $0.34\%$ apparently speaks of the best overall performance which belongs to the mB1WC hybrid -- the only functional that tightly satisfied the $1\%$ criterion for all four considered MAREs. Moreover, the values of MARE$_\text{1}$, MARE$_\text{3}$, and MARE$_\text{T}$ are found to be below $0.5\%$ threshold confirming its perfect suitability for the complete description of the crystalline structure of SrRuO$_3$. 

An analysis of previous study in which the older version of CRYSTAL code was employed [\onlinecite{garcia_80}] reveals that MARE$_\text{T}$ values of DFT results provided therein (LDA, PBE, PBEsol, WC, PBE0, mB1WC) are worse compared to ours on average by about $0.4\%$. We believe that the most probable
source of this discrepancy lies in the basis sets utilized in the calculations and therefore has nothing to do with versions of the CRYSTAL code. Since basis set for Sr is well-tested and identical in both studies, basis sets for O are of the same reliable quality, the factor that remains is the difference between the basis sets for Ru. We have adopted a valence part specially optimized for SrRuO$_3$, and it seems that our choice paid off, even despite the tighter convergence criteria applied in the work of Garc\'{i}a-Fern\'{a}ndez $et$ $al.$ On the other hand, a direct comparison to the previous plane-wave based LDA calculations, carried out by three groups using two distinct $ab$ $initio$ packages [\onlinecite{zayak_77}, \onlinecite{rondinelli_78}, \onlinecite{miao_82}], shows a good agreement with the tendencies of LDA noticed in our study: MARE$_{\text{1}} > 1\%$, MARE$_{\text{2}} > 2\%$, and MARE$_{\text{3}} < 0.5\%$. A somewhat worse overall performance of these works (Zayak $et$ $al.$ with MARE$_\text{T} = 1.11\%$, Rondinelli $et$ $al.$ with MARE$_\text{T} = 1.14\%$, Miao $et$ $al.$ with MARE$_\text{T} = 1.1\%$) could be influenced by different technical parameters, specific selection of $E_{\text{C}}^{\text{LDA}}[n]$ parametrization, and even the way the DFT functionals are implemented in a particular code [\onlinecite{granas_105}].

\subsection{Electronic and magnetic structure}

\begin{figure}
\centering
{
\includegraphics[scale=0.55]{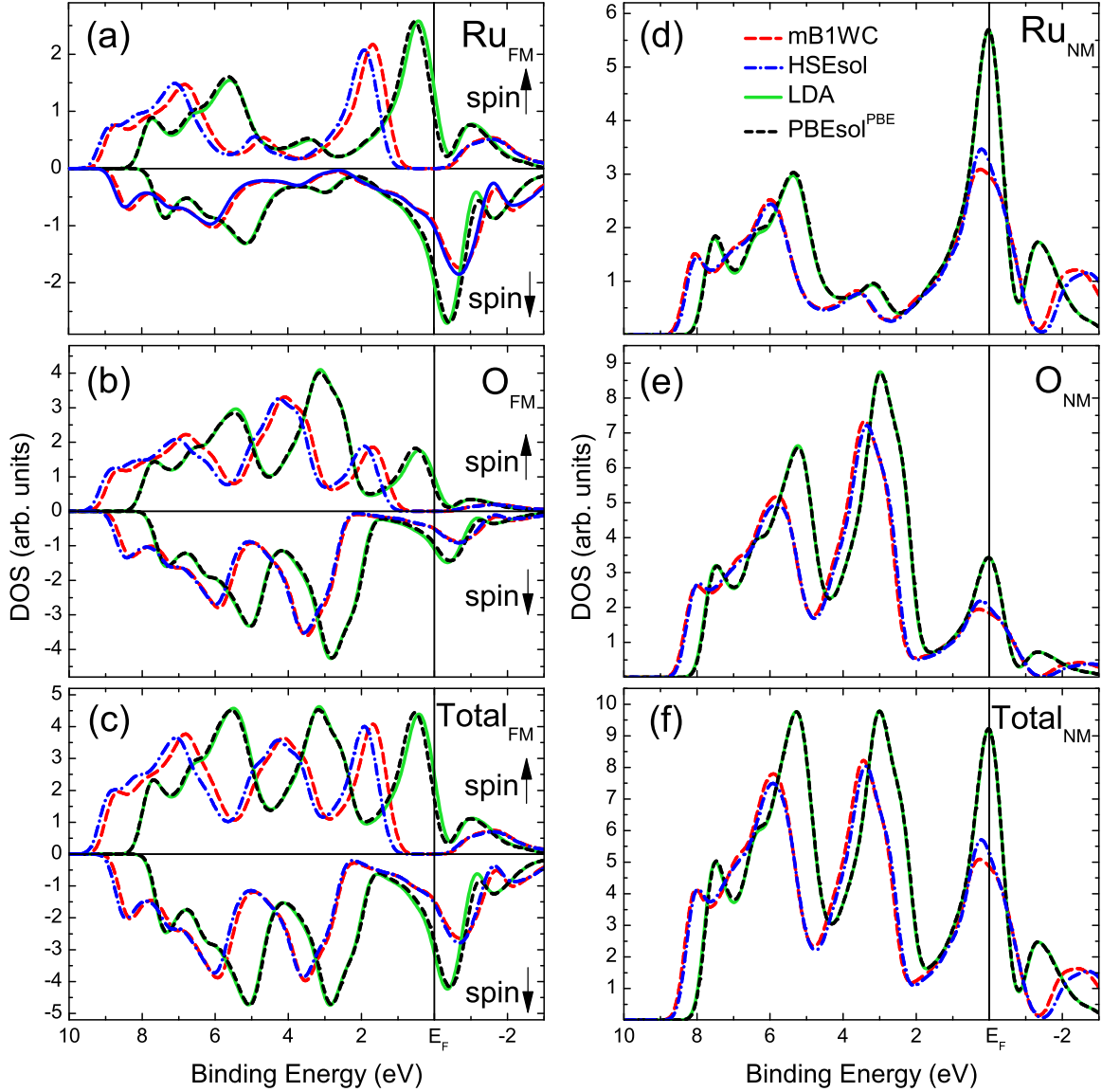}
}
\caption{\label{fig2}
The electronic structure of orthorhombic SrRuO$_3$ evaluated using different DFT approximations: LDA, GGA, and hybrids. (a), (b), and (c) were obtained by performing ferromagnetic (FM), while (d), (e), and (f) non-magnetic (NM) calculations with the experimental geometry at 1.5 K [\onlinecite{bushmeleva_66}]. The Fermi energy ($E_\text{F}$) is set at zero.}
\end{figure}   

The calculated electronic structure and magnetic moment of SrRuO$_3$ can be found in Fig. \ref{fig2} and Table \ref{tab1}, respectively. We have applied the experimental geometry while investigating these properties, since it allows us to eliminate the differences that may rise due to the non-equivalent lattice parameters, distorting the direct comparison between the performance of tested DFT approaches. Our decision was motivated by the work of Gr\r{a}n\"{a}s $et$ $al.$ [\onlinecite{granas_105}] in which the calculated magnetic moments appeared to be quite sensitive to the chosen equilibrium volume of SrRuO$_3$. In addition, we have excluded several functionals from Fig. \ref{fig2} leaving only those that satisfied the $1\%$ of MARE$_\text{T}$ criterion and reflected the typical behaviour of the particular rung of the ladder. Therefore, the remaining functionals represent LDA, GGA (PBEsol\textsuperscript{PBE}), and hybrid approximations with a fraction of HF exchange set to $16\%$ (mB1WC) and $25\%$ (HSEsol). Concerning the electronic structure, the NM density of states (DOS) indicates that valence band of SrRuO$_3$ is formed by strongly hybridized Ru $4d$ and O $2p$ orbitals, with the former ones being dominant in the vicinity of $E_\text{F}$ and the latter ones being more pronounced at higher binding energy regions. The position of the main peaks and shape of the spectra are in a good overall agreement with some recent ultraviolet and x-ray photoemission spectroscopy measurements [\onlinecite{kim_106, toyota_107, grebinskij_108}], and this result only quantitatively depends on the choice of the functional. One can note that performance of LDA and GGA in reproducing NM DOS is identical, while hybrids demonstrate a suppression of electronic states at $E_\text{F}$ and a spectral shift of $\sim$0.5 eV for higher binding energies. However, the picture becomes qualitatively different when energetically more favourable FM solution appears on the scene. In this case, not only does the spectral shift get more pronounced for the mB1WC and HSEsol schemes, but also a band gap opens in the spin-up channel paving the way for the half-metal state in which conduction of the system is assured solely by the spin-down electrons. In the meantime, LDA and GGA frameworks continue to demonstrate metallic behaviour, as their first peak in the spin-up channel clearly crosses $E_\text{F}$. A careful look at the FM part of Fig. \ref{fig2} reveals that LDA exhibits a larger overlap of the spin-up and spind-down bands at $E_\text{F}$ compared to that of GGA, and this is also reflected in the values of magnetic moment in Table \ref{tab1}: $\mu_\text{LDA}=1.63$ versus $\mu_\text{GGA}=1.92-1.96$ $\mu_\text{B}$ per formula unit. If one would take into account experimental uncertainty, $\mu_\text{LDA}$ could practically ideally match the provided result of magnetization measurement in SrRuO$_{3}$. Hybrids, on the other hand, yield a saturated magnetic moment of 2 $\mu_\text{B}$ per formula unit which has not been reported from the laboratories so far. Thus, as long as the experimental observation of half-metallicity and the saturation of magnetic moment remains a serious challenge [\onlinecite{verissimo_81}], LDA seems to be the functional of choice for the realistic reproduction of electronic and magnetic structure of SrRuO$_{3}$. This outcome is also confirmed by different basis set, for instance, plane-waves [\onlinecite{miao_82}] or muffin-tin orbitals [\onlinecite{granas_105}], calculations. 

\subsection{Final remarks}

\begin{figure}
\centering
{
\includegraphics[scale=0.50]{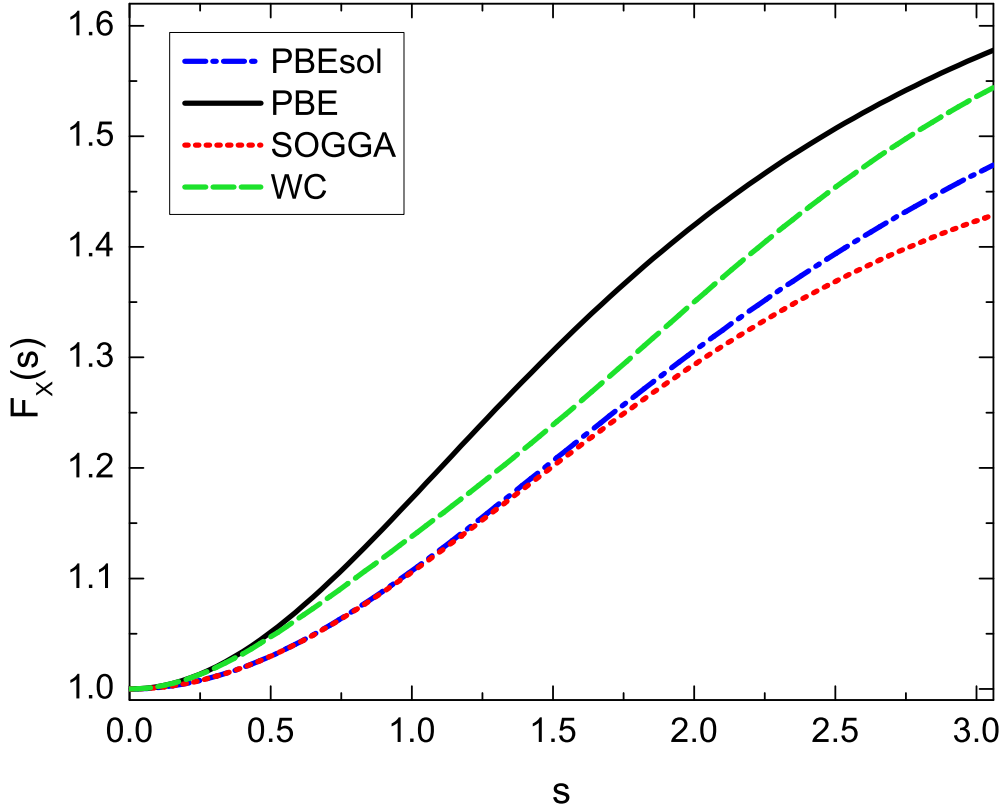}
}
\caption{\label{fig3}
Exchange enhancement factor $F_{\text{X}}(s)$ of different GGAs as the function of the reduced gradient $s$.}
\end{figure}

\begin{figure}
\centering
{
\includegraphics[scale=0.50]{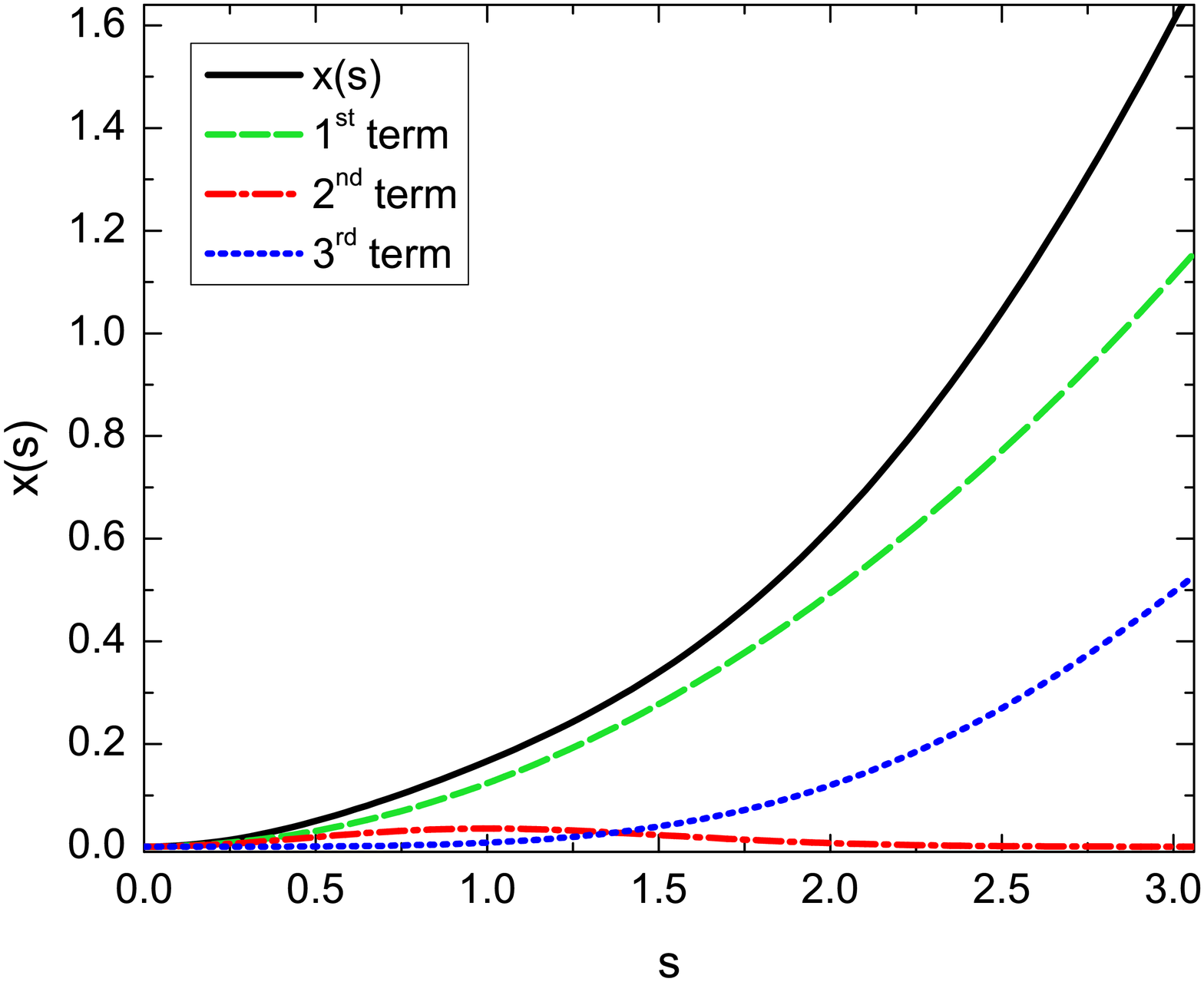}
}
\caption{\label{fig4}
Behaviour of the terms of $x(s)$ [Eq. (\ref{eq:equ10})] introduced in the WC functional.}
\end{figure}

The fact that we used the same correlation functional for PBE, PBEsol\textsuperscript{PBE}, PBE0, SOGGA, and WC allows us to present a few significant insights into the inner structure of SrRuO$_{3}$. Firstly, a simple step that leads from $F^{\text{PBE}}_{\text{X}}(s)$ to $F^{\text{PBEsol}}_{\text{X}}(s)$ appears to be the most important for the accurate description of the geometry of SrRuO$_{3}$. This can be noticed by analysing the overall performance of PBE, PBEsol\textsuperscript{PBE}, and PBE0 (MARE$_\text{T}$ in Table \ref{tab1}) -- the inclusion of a fraction of HF exchange into the PBE framework is not that effective compared to the modification of parameter $\mu$, which was set from $\mu^{\text{PBE}}=0.2195$ to $\mu^{\text{PBEsol}}=10/81\approx 0.1234$. Secondly, the modification of parameter $\beta$ in the correlation part ($\beta^{\text{PBE}}=0.0667\rightarrow \beta^{\text{PBEsol}}=0.046$) has much less impact -- seen from comparison between MARE$_\text{T}$ of PBEsol\textsuperscript{PBE} and PBEsol -- indicating that the magnitude of the exchange energy is substantially larger compared to the correlation energy. Besides, a slightly better performance of the PBEsol\textsuperscript{PBE} combination implies that the restoration of GE for correlation is more relevant for bulk SrRuO$_{3}$ than the original choice to reproduce exchange-correlation energy of jellium surface at the TPSS level [\onlinecite{pbesol_18}]. A close look at Fig. \ref{fig3} reveals that $F^{\text{PBE}}_{\text{X}}(s)$ and $F^{\text{WC}}_{\text{X}}(s)$ are nearly identical for $s\leq 0.5$. Having in mind that MARE$_\text{T}$ results obtained with PBE and WC functionals are in a noticeable discrepancy, one can make an assumption that the average reduced density gradient $s$ should exceed this value in SrRuO$_{3}$. Although the WC approach has rather complicated form, which ensures the same behaviour as PBE for $s \rightarrow 0$ and $s \rightarrow \infty$, examination of the terms in $x(s)$, presented in Fig. \ref{fig4}, shows that at $s$ values larger than $\sim$0.5 the first term $\frac{10}{81}s^2$ starts dominating over the remaining two. Since this term is exactly the same as the one in PBEsol, it becomes clear why these two functionals with distinct $\mu$ values are both able to improve the PBE calculations. This once again confirms the significance of the exact second-order GE for the description of the crystalline structure of SrRuO$_{3}$.

Another close look at Fig. \ref{fig3} indicates that the SOGGA and PBEsol $F_{\text{X}}(s)$ curves are almost identical up to $s\leq1.5$. This observation is perfectly consistent with a very similar performance of SOGGA and PBEsol\textsuperscript{PBE} reflected in the results of all four MAREs. If the average value of $s$ would be higher, we should obtain a more pronounced difference between SOGGA and PBEsol\textsuperscript{PBE}. However, the difference is negligible meaning that the average value of $s$ should fall in the range of $0.5\leq s \leq 1.5$. Such range testifies about a moderately-varying density in SrRuO$_{3}$. It is also interesting that another two similarly performing functionals -- PBEsol and WC -- have different parameters in exchange and correlation parts. As we have already found out that the first term in $x(s)$ makes the largest contribution to $F^{\text{WC}}_{\text{X}}(s)$, one can note that the effect of the addition of remaining two terms somewhat corresponds to the effect of the modification of correlation functional in PBEsol. 

A comparison between the non-screened PBE0 hybrid and its screened HSE06 counterpart also offers a few thoughts on the intrinsic features of SrRuO$_{3}$. On one hand, only a slightly improved performance of the HSE06 functional shows that the non-screened LR part of the HF exchange interaction, present in the PBE0 scheme, is quite unimportant. A beneficial reduction of the self-interaction error, common for GGAs, appears to take place at the SR distance, which turns out to be not so short, extending over two or three chemical bonds in various crystalline materials [\onlinecite{henderson_109}]. On the other hand, a worse performance of the PBE0 functional unveils that the inclusion of the LR exact exchange has a small but negative impact on the geometry of SrRuO$_{3}$. This is not an unexpected trend, though, since the LR portion of the HF exchange may lead to unphysical results in solids [\onlinecite{lucero_110}], especially metallic systems, due to the incomplete cancellation with the approximate correlation functional [\onlinecite{janesko_111}]. However, a factor that has a greater influence on the performance of hybrids is the amount of exact exchange incorporated through the mixing parameter $a$. As we have already found that PBEsol and WC demonstrate similar values of MAREs and HSE-type correction is not essential, a comparison between HSEsol and mB1WC allows to take a look at the effect of $a$ reduction from 0.25 to 0.16. One can note that mB1WC produces noticeably better results which in turn indicate that the typical $25\%$ value of the HF exchange is not the best option for SrRuO$_{3}$. A fraction of $16\%$, in the meantime, is in agreement with the recent observation that a mixing of $10-15\%$ seems to be more appropriate choice for the description of the perovskite family [\onlinecite{franchini_112}]. The advantage of mB1WC over HSEsol also points to the smaller self-interaction error the DFT approaches deal with in this perovskite.

\section{Conclusions}

In this work, we have investigated the crystalline structure of ground-state orthorhombic SrRuO$_3$ by applying a bunch of functionals from three families of DFT approximations: LDA, GGAs, and hybrids. The calculated equilibrium structural parameters -- (a) lattice constants and volume, (b) tilting and rotation angles, and (c) internal angles and bond distances within RuO$_6$ octahedra -- were compared with the low-temperature experimental data. Our analysis of the obtained results allows to highlight several important points. Firstly, the PBE functional is apparently the worst performer among all the contestants, therefore it should be considered as the last option for the geometry optimization of SrRuO$_3$. Secondly, the restoration of the exact second-order density GE for exchange -- in one way or another incorporated into the revised PBEsol, SOGGA, and WC GGAs -- effectively ameliorates performance of the PBE approach, especially for the lattice constants and volume. Thirdly, the inclusion of a portion of the HF exchange also has a positive impact on the crystalline structure of SrRuO$_3$ with respect to the PBE results. In addition, the HSE-type screening of the LR part of the exact exchange is not an essential ingredient for the improvement provided by the hybrid framework. Fourthly and finally, a combination of the revised GGAs with the hybrid scheme seems to be the most appropriate tool for the accurate description of the external [(a)] and internal [(b) and (c)] structural parameters simultaneously. However, amounts of HF exchange smaller than the standard $25\%$ should be preferred. Based on these findings, we are able to offer some recommendations for further theoretical modelling of experimentally observed phases of SrRuO$_3$. If the research is limited to the lattice constants and volume, one should give a chance to the revised GGAs for solids (PBEsol/PBEsol\textsuperscript{PBE}, SOGGA, WC) due to their impressive performance and substantially lower computational cost compared to the hybrids. But if the field of interest comprises interoctahedral angles or full geometry of the system, one should lean towards the hybrids which are based on the revised GGAs (mB1WC, HSEsol). Although the internal geometry of RuO$_6$ is well described by both families of functionals, in case of full geometry of the system hybrids are superior because of their significantly better performance in reproducing tilting and rotation angles. It is also interesting to note that additionally calculated electronic and magnetic structure of SrRuO$_3$ indicates that LDA, despite its striking simplicity, can be treated as the most universal of all the tested functionals. A reasonable geometry optimization followed by a realistic representation of electronic structure and magnetic moment shows that this approximation can be a serious competitor if the crystalline structure is not the only issue the research is focused on.   

\begin{acknowledgments}
The authors are thankful for the HPC resources provided by the ITOAC of Vilnius University. 
\end{acknowledgments} 

\providecommand{\noopsort}[1]{}\providecommand{\singleletter}[1]{#1}%

\end{document}